\def\be{\begin{eqnarray}}
\def\ee{\end{eqnarray}}
\def\ps{p\hspace{-0.075in}/}
\def\pis{\pi\hspace{-0.075in}/}
\def\dels{\partial\hspace{-0.075in}/}
\def\half{{\textstyle{1 \over 2}}}
\def\ihalf{{\textstyle{i \over 2}}}
\def\D{{\cal D}}
\documentstyle[12pt]{article}
\begin{document}
\bigskip
\Large
\begin{center}
\bf{From The Superparticle Path Integral To Superfield Theory}\\

\bigskip

\normalsize
by\\
\bigskip

J.Grundberg\\
{\it Department of Theoretical Physics\\
The Royal Institute of Technology\\
S-100 44 Stockholm\\
SWEDEN\\
\bigskip
and\\}
\bigskip
U.Lindstr\"om and H.Nordstr\"om\\
{\it ITP\\
University of Stockholm\\
Vanadisv\"agen 9\\
S-113 46 Stockholm\\
SWEDEN}\\
\end{center}
\vspace{1.0cm}
\normalsize
{\bf Abstract:} We investigate the hitherto unexplored relation between the
superparticle path integral and superfield theory. Requiring that the path
integral
has the  global symmetries of the classical action and obeys the natural
composition
property of path integrals, and also that the discretized action has the
correct
naive continuum limit, we find a viable discretization of the (D=3,N=2)
free superparticle action. The resulting propagator is not the usual superfield
one.
We extend the discretization to include the coupling to an external gauge
supermultiplet and use this to show the equivalence to superfield theory.
This is
possible since we are able to reformulate the superfield perturbation theory
in terms
of our new propagator.

\eject

\begin{flushleft}
{\bf Introduction}
\end{flushleft}
\bigskip
There have been numerous attempts to quantize both the
massive and the massless superparticle \cite{EVAN1}-\cite{MIKO}. Both the
massless and massive models are invariant under a certain fermionic symmetry;
the
Siegel symmetry \cite{SIEG1}. For the massive case
quantization has been carried out both using BV-BRST methods \cite{GREE} and
using covariant methods \cite{EVAN1}. In the massless case a covariant
separation of the models first and second class constraints is not possible in
general \cite{BENG}. Attempts to circumvent this problem have been
made using BV-BRST methods \cite{LIND1} and using harmonic
superspace methods \cite{NISSI}. Also non-covariant quantization has been
described \cite{EVAN2}. A constructive path integral quantization has, to our
knowledge, only been attempted in \cite{MIKO}, however.

We have given a brief
report on the definition of the path integral in a letter \cite{GRUN}.
Here we present
the calculation in more detail along with an extensive discussion of the
coupling to a background field.

Our construction starts from a set of (natural)
requirements on the path integral; it should have the global symmetries of the
classical action and it should obey the usual composition property of a path
integral. In addition, we require the discretized action to have the correct
naive continuum limit (i.e., assuming that the difference between the values
of a
function at time $t$ and $t+\varepsilon$ is of order $\varepsilon$).
 We find a discretization that complies with these
demands and construct the propagator. This propagator differs from the usual
field theory propagator as well as from the propagator derived in \cite{MIKO}.
In particular, it contains an additional factor of an inverse momentum squared.
This factor is dictated on dimensional grounds from the composition property
and the dimension of the measure. Since it is unclear what physical meaning to
ascribe to the superspace propagator for the free theory, we study the coupling
to a background gauge multiplet and find that the coupling is directly to the
gauge potentials $\cal A$ rather than to the gauge prepotentials, which is the
case for
the superfield theory. We derive the Feynman rules and compare the perturbation
expansion of the effective action to that of a (massless) chiral superfield
coupled to
a gauge super multiplet. We show equivalence explicitly to second order in the
external field. In the process
of doing this we show that it is possible to organize the superfield supergraph
calculations in such a way that the propagator agrees with the one we derive
from the superparticle. We have verified that the equivalence in fact holds
to all
orders in $\cal A$.

As a further check on our methods, a similar
calculation has been carried out comparing the superparticle in a light cone
gauge and the light cone superfield theory \cite{NORD}.

Our analysis highlights that the path integral is a formal object
which has to be given content by some evaluation prescription. This is
particularly the case for fermionic variables. We also want to draw
attention to the fact that the relation between the first quantized theory
and the field theory is not always as simple as in the ordinary scalar field
case. This is worth having in mind when trying to find a string field theory.

The organization of the article is as follows: In Sec.{\bf II} we recapitulate
the tensor calculus of \cite{GAUN} for constructing Siegel invariant actions
in $D=3$ and present the action for the free superparticle as well as the
action for a superparticle coupled to a gauge super multiplet. In Sec. {\bf
III} we exhibit a Legendre transform of the abovementioned actions and give
the corresponding phase space actions which we subsequently use in the path
integral. Sec. {\bf IV} is devoted to the definition of the propagator via a
discretization of the path integral. In Sec. {\bf V} a survey of the
supergraph rules for chiral fields coupled to a gauge prepotential superfield
can be found along with the discussion of how to reorganize the perturbation
expansion using the propagator derived from our particle path integral. A
comparison of the superfield theory vertices to the interaction parts
of our particle coupled to a gauge potential is then given in Sec. {\bf VI}.
Our conclusions constitute Sec. {\bf VII}. In the {\bf Appendix} we show the
equivalence between the field theory and particle theory to all orders in the
background field.

\bigskip
\begin{flushleft}
{\bf II. Siegel Invariant Actions}
\end{flushleft}
\bigskip

In \cite{GAUN} a tensor calculus for the Siegel-symmetry of $D=2$ and $3$
superparticles is presented. It is based on a reformulation of the theories as
supersymmetric $\sigma$-models that are invariant under local world-line
superconformal transformations. The $N=2, D=3$ massive superparticle is
described as follows:

The world line is extended to a $N=2$ world line in superspace with coordinates
$\{z^M\}=(t,\eta , \bar \eta)$ and the $D=3, N=2$ flat target superspace is
coordinatized by  $(X^\mu (z), \Theta  _\alpha (z), \bar \Theta ^\alpha
(z))$, a space-time vector and
two space-time spinors, all of which are world-line scalar fields. These fields
are subject to the constraints
\be
&&\bar{D}X^\mu = -\ihalf (\bar{D}\bar \Theta )
\Gamma ^\mu \Theta \quad and \quad c.c\cr
&&\bar{D}\Theta=0 \quad and \quad c.c.,\label{cons}
\ee
where
\be
\left\{{\Gamma ^\mu ,\Gamma ^\nu }\right\}=-2\eta ^{\mu \nu },
\qquad \eta^{\mu \nu}=diag(-++)
\ee
and
\be
D \equiv \partial _\eta +i \bar{\eta}\partial _t,\qquad
\bar{D}\equiv -\partial _{\bar{\eta}} -i \eta\partial _t.
\ee
The constraints (\ref{cons}) lead to the following component expansion:
\be
\Theta _\alpha \mid =\theta _\alpha &\qquad&\bar{\Theta}^\alpha \mid
=\bar{\theta}^\alpha\cr
D\Theta _\alpha \mid=\lambda _\alpha &\qquad& \bar{D}\bar{\Theta} ^\alpha
\mid=-\bar{\lambda} ^\alpha \cr
\half [\bar{D},D]\Theta_\alpha \mid =-i\dot{\theta}_\alpha &\qquad&
\half [D,\bar{D}]\bar{\Theta}^\alpha \mid =-i\dot{\bar{\theta}}^\alpha
\label{Psic}
\ee
and
\be
X^\mu \mid &=&x^\mu\cr
DX^\mu \mid &=& -\ihalf \bar{\theta}\Gamma ^\mu \lambda\cr
\bar{D}X^\mu \mid &=& \ihalf \bar{\lambda}\Gamma ^\mu \theta, \label{Xcom}
\ee
where $\mid$ denotes the $\eta =0$ projection. From (\ref{cons}) we also find
the identity
\be
\bar{\lambda}^\alpha \lambda _\beta = \sqrt{-\pi^2}\delta^\alpha
_\beta +(\Gamma ^\mu)^\alpha _\beta \pi _\mu \label{iden}
\ee
where
\be
\pi ^\mu \equiv \dot{x}^\mu +\ihalf \left({\dot{\bar\theta}\Gamma ^\mu \theta
-\bar \theta \Gamma ^\mu \dot{\theta}}\right).
\ee
The constraints are invariant under $N=2,\quad D=3$ space-time
supersymmetry:
\be
\delta X^\mu&=&\ihalf \left({\bar{\rho}\Gamma ^\mu \Theta - \bar\Theta \Gamma
^\mu \rho}\right)\cr
\delta \Theta&=&\rho , \qquad \delta \bar \Theta = \bar \rho
\ee
with $\rho$ and $\bar\rho$ two constant anticommuting space-time spinors.
Furthermore there are also the local $N=2$ superconformal transformations
\be
\tilde{\delta}z^M=B\dot{z}^M+\ihalf \bar{D}BDz^M+\ihalf DB\bar{D}z^M
\label{vart} \ee
with
\be
B\mid=-b(t),\qquad DB\mid =2i\bar{\varepsilon} (t),\qquad\bar{D}B \mid
=-2i\varepsilon (t).
\ee
The action induced by (\ref{vart}) on the components in (\ref{Psic},\ref{Xcom})
show that $b(t)$ is the parameter for infinitesimal reparametrizations while
the identifications
\be
\varepsilon \equiv {{\bar{\lambda}\kappa} \over {\bar{\lambda}\lambda}},
\ee
yield
\be
\delta _\kappa \theta _\alpha &=& \half
\left({\delta_\alpha^\beta+{{\pis_\alpha^\beta}\over{\sqrt{(-\pi
^2)}}}}\right)\kappa_\beta\cr
\delta _\kappa \bar\theta ^\alpha &=& \half \bar\kappa^\beta
\left({\delta^\alpha_\beta+{{\pis^\alpha_\beta}\over{\sqrt{(-\pi
^2)}}}}\right)\cr
\delta _\kappa x^\mu &=& \ihalf\left({\bar{\theta}\Gamma ^\mu \delta
_\kappa \theta-
\delta_\kappa\bar{\theta}\Gamma ^\mu  \theta}\right),
\ee
i.e., the Siegel transformations, \cite{SIEG1}. A
covariantization of the theory with respect to the transformations (\ref{vart})
proceeds by introducing covariant derivatives on scalar fields as
\be
\nabla \Phi \equiv E^{-1}D\Phi, \qquad \bar{\nabla} \Phi \equiv
E^{-1}\bar{D}\Phi
\ee
with
\be
E\equiv \sqrt{-\bar{D}\bar{\Theta}D\Theta}
\ee
and covariant time derivative
\be
\nabla _0\Phi\equiv \ihalf \{\nabla,\bar{\nabla}\}\Phi.
\ee
The world-line superconformal transformations are
\be
\delta(\nabla \Phi)=iL\nabla \Phi, \qquad \delta(\bar{\nabla}
\Phi)=-iL\bar{\nabla} \Phi
\ee
where
\be
L=\bar{L}\equiv -{\textstyle{1 \over 4}}[D,\bar{D}]B.
\ee
A general Siegel-invariant action may now be written as
\be
S=\int\limits_0^1{d^3z{{\cal L}}\left({X,\Theta,\bar{\Theta},\nabla \Theta,
\bar{\nabla}\Theta, \nabla _0\Theta,...}\right)}. \label{gact}
\ee
The usual (second order) $N=2$ massive superparticle action \cite{GAUN2}
results from
the choice
\be
{{\cal L}}=-{{m}\over 2}\bar{\Theta}\Theta .\label{fpla}
\ee
After integrating out the $\eta$-dependence,
\be
S&=&-{{m}\over 2}\int\limits_0^1{d^3z\bar{\Theta}\Theta}\cr
&=&-m\int\limits_0^1{d\tau \left\{{\sqrt{-\pi
^2}+\ihalf \left({
\dot{\bar\theta}\theta-\bar\theta\dot\theta}\right)}\right\}}.\label{pact} \ee

Up to this point the description of the covariantization has followed
Gauntlett \cite{GAUN} exactly. Now we adapt his results to our needs.

A special case of the general action (\ref{gact}) is
\be
S&=&\int\limits_0^1{d^3z{\cal L}(X,\Theta,\bar{\Theta})}\cr
&=&\half \int\limits_0^1{d\tau [\bar{D},D]{\cal
L}(X,\Theta,\bar{\Theta})}\mid.\label{sac} \ee
Using the chain rule for differentiation, the definition of components
(\ref{Psic},\ref{Xcom}) as well as the constraint relations (\ref{cons}), we
expand the action (\ref{sac}) and find
\be
S=\int\limits_0^1{d\tau \left\{{\half [\lambda_\alpha\D^\alpha
,\bar{\lambda}^\beta \bar\D_\beta ]{\cal L}
+i\dot{\bar{\theta}}^\alpha \bar\D_\alpha {\cal L}
-i\dot{\theta}_\alpha \D^\alpha {\cal L}}\right\}},
\ee
where
\be
\D^\alpha &\equiv & {\partial \over {\partial
\theta_\alpha}}-\ihalf(\bar{\theta}\Gamma ^\mu)^\alpha {\partial \over
{\partial
x^\mu}},\cr
\bar\D_\alpha &\equiv & {\partial \over {\partial
\bar{\theta}^\alpha}}-\ihalf(\Gamma ^\mu \theta)_\alpha {\partial \over
{\partial x^\mu}},
\ee
which implies
\be
\left\{{\bar\D_\alpha , \D^\beta}\right\}=-i(\Gamma ^\mu)_\alpha
^\beta {\partial \over{\partial x^\mu}}.
\ee
Hence $\D_\alpha$ satisfy the global $D=3,\quad N=2$ supersymmetry
algebra (in complex notation).

Using the identity (\ref{iden}), we finally have
\be
S=\int\limits_0^1{d\tau \left\{{\sqrt{-\pi^2} \D^\alpha \bar{\D}_\alpha
{{\cal L}}
+\half\pi_\mu[\D,\Gamma^\mu\bar{\D}]{{\cal L}}
+i\dot{\bar{\theta}}^\alpha \bar{\D}_\alpha {{\cal L}}
-i\dot{\theta}_\alpha \D^\alpha {{\cal L}}}\right\}}.\label{tact}
\ee

If we let ${\cal L}=V(x,\theta ,\bar\theta)$ be a general real scalar
superfield,
this looks very much like a coupling of the particle to a gauge multiplet:
\be
S=\int\limits_0^1{d\tau \left[{ -\sqrt{-\pi^2} {\cal A}+\pi ^\mu {\cal A}_\mu
-i\dot{\theta}_\alpha {\cal A}^\alpha +i\dot{\bar{\theta}}^\alpha \bar{\cal
A}_\alpha}\right]} \label{iact}
\ee
where the gauge potentials are given in terms of the prepotential $V$ trough
\be
\bar{\cal A}_\alpha &=& \bar\D_\alpha V, \qquad
{\cal A}^\alpha = \D^\alpha V,\cr
{\cal A}^\mu &=& \half\left[{\D ,\Gamma ^\mu\bar\D }\right]
V,\qquad
{\cal A} \equiv -\D^\alpha \bar\D_\alpha V.\label{Adef}
\ee
That this describes the $N=2$,$D=3$ vector multiplet is perhaps most easily
seen
by dimensional reduction of the $N=1$,$D=4$ vector multiplet.
${\cal A}^\mu$,$\bar{\cal A}_\alpha$ and ${\cal A}^\alpha$ should certainly be
present and ${\cal A}$ is the reduction of the fourth component of the vector
potential. The constraints imply that they can be expressed in terms of a
real scalar superfield as above. Gauge transformations take the form \be
\delta V = i\left({\Lambda -\bar{\Lambda}}\right), \quad \D^\alpha
\bar\Lambda =\bar\D_\alpha\Lambda =0. \label{Vgtf}
\ee
Using the identity
\be
{{d\Phi} \over {d\tau }}=\pi ^\mu \partial _\mu\Phi +\dot{\bar{\theta}}^\alpha
\bar\D_\alpha \Phi +\dot{\theta}_\alpha \D^\alpha \Phi \label{rela}
\ee
(valid for a general superfield $\Phi$), we find
\be
\delta S=\int\limits_0^1{d\tau {d \over{d\tau}}(\Lambda -\bar\Lambda)},
\ee
so the action (\ref{tact}) transforms as it should under a gauge
transformation.

\bigskip
\begin{flushleft}
{\bf III. The phase space action}
\end{flushleft} \bigskip

As it stands, the action (\ref{tact}) has the disadvantage of involving a
square
root. We can get rid of this by making a Legendre transform with respect to
$\pi
_\mu$. Let
\be
&&L(\pi ^\mu ,\dot{\theta},\dot{\bar{\theta}},x,\theta ,\bar{\theta})\cr
&=&\left\{{\sqrt{-\pi ^2} \D^\alpha \bar\D_\alpha +\half [{\cal
D}^\alpha,(\pis )_\alpha ^\beta \bar\D_\beta
]-i\dot{\bar{\theta}}^\alpha\bar\D_\alpha +
i\dot{{\theta}}_\alpha \D^\alpha }\right\}{\cal L}(x,\theta
,\bar{\theta})\cr
&&.\label{Lagr}
\ee
and define
\be
p_\mu&=&{{\partial { L}} \over {\partial \pi ^\mu}}\cr
H(p,\dot{\theta},\dot{\bar{\theta}},x,\theta ,\bar{\theta})&=&p_\mu \pi ^\mu
-L.
\label{Hdef}
\ee
(The reader is invited to check that this is the usual transition to the
hamiltonian for the case when $L$ does not depend explicitly on $\dot{\theta}$
and
$\dot{\bar{\theta}}$.) Clearly terms in $L$ that are homogeneous of degree
one in $\pi ^\mu$ will not contribute to $H$.We thus find:
\be
H=\left({-i\dot{\bar{\theta}}^\alpha \bar\D_\alpha
+i\dot{{\theta}}_\alpha \D^\alpha}\right){\cal L}. \label{hami}
\ee
Since
\be
p_\mu ={{\partial {L}} \over {\partial \pi ^\mu}}=\left({
{{-\pi _\mu} \over {\sqrt{ -\pi ^2}}}\D^\alpha \bar\D_\alpha +\half
[\D^\alpha ,(\Gamma _\mu )_\alpha ^\beta \bar\D_\beta ]}\right){\cal
L} \ee
we also have a primary constraint
\be
\left[{p_\mu -\half[\D^\alpha ,(\Gamma _\mu )_\alpha ^\beta \bar{\cal
D}_\beta ]{\cal L}}\right]^2=-\left({\D^\alpha \bar\D_\alpha {\cal
L}}\right)^2.\label{pcon}
\ee
Incorporating this into the action with a Lagrange multiplier we get the
phase space action
\be
S^{PS}&=&\int\limits_0^1{d\tau \left\{{p_\mu \pi ^\mu
-e\left[{\left[{p_\mu -\half[{\cal
D}^\alpha ,(\Gamma _\mu )_\alpha ^\beta \bar\D_\beta ]{\cal
L}}\right]^2+\left({\D^\alpha \bar\D_\alpha {\cal
L}}\right)^2}\right]}\right.}\cr
&&\left.{+i\dot{\bar{\theta}}^\alpha\bar\D_\alpha
{\cal L}-i\dot{\theta}_\alpha\D^\alpha {\cal L}}\right\}.\label{pspa}
\ee
To get the final form of the action we write ${\cal L}=-{\textstyle {m\over
2}}\bar\theta\theta +V$ and shift $p_\mu - {\cal A}_\mu \to p_\mu$. The
result is \be
S=\int\limits_0^1{d\tau  \left\{{p_\mu \pi^\mu -e(p^2+(m+{\cal A})^2)-\ihalf
m(\dot{\bar{\theta}} \theta  -\dot{\theta}
\bar{\theta})}\right.}\cr
\left.{+\pi_\mu{\cal A}^\mu +i\dot{\bar\theta}^\alpha \bar{\cal
A}_\alpha -i\theta _\alpha {\cal A}^\alpha }\right\}.\label{sup1}
\ee
This should be compared to the $D=10$ superparticle in a super-Maxwell
background discussed in \cite{ROCE}.

\bigskip
\begin{flushleft}
{\bf IV. The Propagator}
\end{flushleft}
\bigskip

The basic object in a path integral quantization is the propagator
\cite{FEYN}. In the
case of our free superparticle, it is symbolically given by
\be
K(x _f,\theta _f,\bar \theta _f;x _i,\theta _i,\bar \theta _i)=\int{{\it
DeDpDxD\theta D\bar \theta}exp \left\{{iS}\right\}},\label{Kdef}
\ee
where the action $S$ is (\ref{sup1}) with the gauge potentials set to zero.
We follow the usual treatment \cite{POLY} in making the gauge choice
\be
\dot e = 0 \quad \Rightarrow \quad e=T.
\ee
The relation (\ref{Kdef}) is then replaced by
\be
K=\int \limits _0^\infty {d\tau G},
\label{intG}
\ee
where
\be
G(x _f,\theta _f,\bar \theta _f;x _i,\theta _i,\bar \theta _i;T)
=\int{{\it DpDxD\theta D\bar \theta}exp\left\{ iS \right\}}\label{G}
\ee
is a function of the final and initial superspace positions and
\be
S=\int\limits_0^T {d\tau \left\{ {p_\mu \left[ {\dot x^\mu +{\textstyle{i
\over 2}}\dot{\bar \theta }\Gamma ^\mu \theta -{\textstyle{i \over 2}}\bar
\theta
\Gamma ^\mu \dot \theta } \right]-{\textstyle{i \over 2}}m(\dot{\bar \theta}
\theta -\bar \theta \dot \theta )-(p^2+m^2)} \right\}}.
\label{action2}
\ee
It remains to give a meaning to this as yet symbolic expression for $G$. One
way of defining a path integral is to discretize, i.e., to represent a path by
the positions at a finite number of intermediate times, replace the action by a
discretized version and integrate over the intermediate positions. A natural
requirement is that the discretized action should have the original one as the
(naive) limit as the number of time steps goes to infinity, but, as is well
known, different discretizations complying with this condition can still give
different results for the path integral. The root of the difficulty is that one
is effectively summing over non-differentiable or even discontinuous paths, so
the naive continuum limit is naive indeed. (This is the place where
ordering problems enter the path integral scheme.) In addition to this
difficulty, one also has to determine the measure for the integration over
intermediate positions. We will base our treatment of these points on the
symmetries of the problem and a general composition property of path integrals.

The composition property of path integrals is the intuitive rule that one can
calculate the amplitude from $A$ to $C$ by multiplying the amplitude from $A$
to $B$ with the one from $B$ to $C$ and sum over the intermediate positions
$B$ \cite{FEYN}. In symbols
\be
G(3;1;T_1+T_2)=\int{d^3x_2d^2\theta _2d^2 \bar \theta _2G(3;2;T_2)G(2;1;T_1)}
\label{cmp}
\ee
(where the arguments have been abbreviated, e.g.,
$(x_3,\theta_3,\bar{\theta}_3) \to 3$ et.c.). In formal derivations of path
integrals this property expresses the completeness of intermediate states, from
the intuitive point of view it is almost the defining property of path
integrals. In perturbative calculations it is this property that gives the
structure "{\it propagator $\times$ vertex $\times$ propagator $\times$...}", a
structure familiar from field theory and one we would also expect from first
quantized treatment of the superparticle. Finally we note that the
composition property (\ref{cmp}) determines the dimension of $G$. Compared to
the scalar particle path integral, the dimension of $G$ must be such that it
cancels the dimension of the fermionic measure $d^2 \theta d^2 \bar{\theta}$.

For a free superparticle $G(2;1;T)$ is to a large extent determined by the
composition rule (\ref{cmp}) and the symmetries of the problem. The latter,
global $N=2$ Poincar\'e supersymmetry, implies that
\be
G(2;1;T)&=&\int{{{d^3p}\over {(2\pi)^3}}\tilde{G}(p,\theta_2
-\theta_1,\bar\theta_2-\bar\theta_1) }\cr
&\times &exp\left\{{i\left[{p_\mu \left({x^\mu _2-x_1^\mu+{\textstyle{i \over
4}}(\bar\theta_2-\bar\theta_1)\Gamma^\mu
(\theta_2+\theta_1)}\right.}\right.}\right.\cr
&&-\left.{{\textstyle{i \over4}}(\bar\theta_2+\bar\theta_1)\Gamma^\mu
(\theta_2-\theta_1)}\right)-(p^2+m^2)T\cr
&&\left.{\left.{-{{im}\over 4}
\left({(\bar\theta_2-\bar\theta_1)(\theta_2+\theta_1)
-(\bar\theta_2+\bar\theta_1)(\theta_2-\theta_1)}\right)
}\right]}\right\},\label{G2}
\ee
where we have also used the "dynamical" relation
$i\partial_TG=((-i\partial_x)^2+m^2)G$. The translational invariance of
(\ref{G2})
is manifest, the supertranslational one can be expressed as
\be
({\cal Q}^\alpha _2(1) +{\cal Q}^\alpha _1(-1))G = (\bar{\cal Q}_{\alpha 2}(1)
+\bar{\cal Q}_{\alpha 1}(-1))G=0, \label{achi1}
\ee
where
\be
{\cal Q}^\alpha (Z) &\equiv & {\partial \over {\partial \theta _\alpha}}
+\ihalf (\bar \theta \Gamma ^\mu )^\alpha{\partial \over {\partial
x^\mu}}+{{Zm}
\over 2}\bar\theta^\alpha\cr
\bar{\cal Q}_\alpha (Z) &\equiv & {\partial \over {\partial \bar \theta
^\alpha}}
+\ihalf ( \Gamma ^\mu \theta )_\alpha{\partial \over {\partial x^\mu}}+{{mZ}
\over
2}\theta_\alpha\cr
\left\{{{\cal Q}^\alpha (Z) ,\bar{\cal Q}_\beta (Z) }\right\}&=&
i\left({\Gamma^\mu
}\right)^\alpha _\beta {\partial \over {\partial x^\mu}}+m\delta ^\alpha
_\beta Z
\cr
\left\{{
{\cal Q}^\alpha (Z) ,{\cal Q}^\beta (Z)
}\right\}&=&
\left\{{\bar{\cal Q}_\alpha (Z) ,\bar{\cal Q}_\beta (Z) }\right\}=0
\ee
are the generators of $N=2$ supersymmetry with central charge proportional
to $Z$.

At this point we could try to impose the composition rule (\ref{cmp}) to
determine $\tilde G$. (We note in passing that for a scalar particle this
would indeed give $\tilde G=1$.) The result would not be unique, however.
Instead we note that the symmetries of the problem do not prevent us from
imposing the antichirality condition
\be
\D^\alpha _2 (1) G=0, \label{achi}
\ee
where
\be
\D^\alpha (Z) &\equiv & {\partial \over {\partial \theta _\alpha}}
-\ihalf (\bar \theta \Gamma ^\mu ){\partial \over {\partial x^\mu}}-{{Zm} \over
2}\bar\theta^\alpha\cr
\bar\D_\alpha (Z)&\equiv & {\partial \over {\partial \bar \theta ^\alpha}}
-\ihalf ( \Gamma ^\mu \theta ){\partial \over {\partial x^\mu}}-{{Zm} \over
2}\theta_\alpha\cr
\left\{{\D^\alpha (Z),\bar\D_\beta (Z)}\right\}&=&-i\left({\Gamma^\mu
}\right)^\alpha _\beta {\partial \over {\partial x^\mu}}-Zm\delta ^\alpha
_\beta
\cr
\left\{{\D^\alpha (Z) ,\D^\beta (Z)}\right\}&=&\left\{{\bar\D_\alpha (Z)
,\bar\D_\beta (Z)}\right\}=0
\ee
are the $N=2$ supercovariant derivatives with central charge proportional
to $Z$.

The antichirality condition (\ref{achi})
fixes the  $\theta,\bar\theta$ dependence of $\tilde G$ and adding the
composition condition  (\ref{cmp}) we find
\be G(2;1;T)&=&\int{
{{d^3p}\over{(2\pi)^3}}{1 \over {p^2 +m^2}}}exp\left\{{i\left({ p_\mu \left[{
x ^\mu _2-x ^\mu _1
}\right.
}\right.
}\right.
\cr
&-&\left.{{\textstyle{i \over 2}}\bar \theta _2
\Gamma ^\mu(\theta _2-\theta _1) +{\textstyle{i \over 2}}(\bar \theta _2 -\bar
\theta _1)\Gamma ^\mu\theta _1 }\right]\cr
&-&\left.{\left.{
(p^2+m^2)T-{\textstyle{i \over2}}m
\left[{
(\bar \theta _2 -\bar \theta _1)\theta _1-\bar \theta _2(\theta
_2-\theta _1)
}\right]}\right)
}\right\}. \label{disc}
\ee
This is the final result for the free superparticle propagator. Apart from the
symmetries, path integral properties were used in the derivation
in the form of the dynamical relation $i\partial _\tau G=((-i\partial _x)^2
+m^2)G$ and the composition property (\ref{cmp}).

 We can now iterate
(\ref{disc}) to build up a path integral. The result is
\be
G(f;i;T)&=&\cr
&&\mathop{lim}\limits_{N\to \infty}\int {\prod_{k=1}^{N-1} {\left(
{{{d^3p_kd^3x_kd^2\theta _kd^2\bar \theta _k} \over {(2\pi )^3(p_k^2+m^2)}}}
\right){{d^3p_N} \over {(2\pi )^3(p_N^2+m^2)}} }}
exp\left\{{iS}\right\},\cr
&.&
\ee
where
\be
S&=&\sum\limits_{k=1}^N {\left({-(p_k^2+m^2)\varepsilon
+p_\mu ^k\pi ^\mu_{k,k-1}\varepsilon}\right.}\cr
&&-\left.{{\textstyle{i \over 2}}m\left[ {(\bar \theta _k-\bar \theta
_{k-1})\theta _{k-1}-\bar \theta _k(\theta _k-\theta _{k-1})}
\right]}\right)\label{Fdis}
\ee
and
\be
\pi ^\mu _{k,k-1}\varepsilon\equiv x_k^\mu -x_{k-1}^\mu -{\textstyle{i
\over 2}}\bar \theta _k\Gamma ^\mu (\theta _k-\theta _{k-1})+{\textstyle{i
\over
2}}(\bar \theta _k-\bar \theta _{k-1})\Gamma ^\mu \theta _{k-1}.
\ee
We have thus determined the measure and the discretized action. Note that the
latter has the original Siegel-invariant action as the naive continuum limit.
Note also the asymmetry in $\theta$ and $\bar \theta$. We remind the reader
that a similar asymmetry exists in the path integral approach to fermion
fields using coherent states \cite{OHNU}.

The result (\ref{disc}) for the propagator was not quite unique. In addition to
symmetries and general path integral ideas, (composition rule plus dynamical
equation), we imposed an antichirality condition. It is clear that we could
have used chirality instead, and the only difference would be $\bar \theta
\leftrightarrow \theta$ everywhere. Are there other possibilities?

Some insight into this question is obtained by noting that the result for $G$
can be written
\be
&&G(f;i;T)=\cr
&&{\textstyle{1 \over
2}}\D_f^\alpha\D_f^\beta \bar\D_\beta ^i\bar\D_\alpha
^i\left[{\delta^4(\theta _f-\theta _i)\int{{d^3p \over {(2\pi)^3}}{1 \over
{p^2+m^2}}e^{\left\{{-iT(p^2+m^2)+ip_\mu (x_f^\mu -x_i^\mu
)}\right\}}}}\right],\cr
&&.\label{Ddelta}
\ee
where we have suppressed the values of the central charge.
Since $(\bar \D^i(-1))^2$ acts on  $\delta ^4(\theta_f
-\theta_i)exp\left\{{ip_f(x_f-x_i)}\right\}$ we can replace it by $(\bar {\cal
D}_f(1))^2$, and, combining $\D^2\bar\D^2$ with the $(p^2+m^2)^{-1}$,
we have the antichiral projection operator $\D^2\bar\D^2 /
(p^2+m^2)$ acting. Some thought shows that a projection operator acting on
$\delta ^4(\theta_f -\theta_i)$ is indeed what is needed to satisfy the
composition property (\ref{cmp}). In addition to the antichiral and chiral
projection operators we may also consider the linear projector as a third
and final independent projection operator. The discretized action
corresponding to this alternative contains extra $\theta$-terms that vanish
in the continuum limit. We will
comment further on this in the conclusions.

As was mentioned in the introduction, another attempt to define the
superparticle path
integral by discretization has been made in \cite{MIKO}. Those authors
proceed by writing down a discretization compatible with the global
supersymmetry. Their discretization is the one that is obtained by using the
"midpoint rule"
\be
x(t) &\to& {{x_{i+1}+x_i}\over 2}\cr
\dot{x}(t) &\to& {{x_{i+1}-x_i}\over \varepsilon}\label{midp}
\ee
for both bosonic and fermionic variables. In the notation of this paper it
corresponds to having $\tilde G$ independent of $\theta ,\bar{\theta}$. Such
a discretization does not satisfy the composition property, but has the
advantage of
being always applicable. As is noted in  \cite{MIKO}, the result depends on
whether the
number of steps in the discretization is even or odd. The authors of
\cite{MIKO}
choose an odd number of steps to ensure an even number of integrals for each
Lorentz
component of $\theta$. Their final result for $K$ (denoted $G$ in
\cite{MIKO}), for a massless superparticle is \be
K=\D^{N'}\int{{{d^Dp}\over{(2\pi )^D}}{exp
{\left\{{ip(x_f-x_i)}\right\}}\over {p^2}}\delta^{N'}(\theta
_f-\theta_i)} \ee
where $\D^{N'}$ is the antisymmetrized product of all the covariant
derivatives. This differs from our result in two respects, in the
derivative structure and in the momentum dependence. Our result for $G$
implies that  $K$ has  $p^4$ in the denominator instead of the $p^2$ above.
On the other hand the authors of \cite{MIKO} associate factors $p^{-2}$ with
interaction over $\bar\theta _k,\theta_k$ in the discretization. Our
objection to this approach is that it violates the basic composition rule of
path integrals, and as a consequence it is harder to understand how to
incorporate interactions. If this is done trough a discretization, one would
certainly also need to use the propagator obtained with an even number of
intermediate steps and thus obtain a perturbation theory with two kinds of
propagators. A more techniqual objection is that to associate the $p^{-2}$
factors in the measure with the $\theta , \bar\theta$ integration rather
than the $p$ integration seems unnatural to us. The discretization is a set
of points in superspace $(x,\theta ,\bar\theta )$ with $p$ associated with
the links between these points, and one would not expect these two kinds of
variables to mix in the measure for a free particle.

What about the relation to superfield theory? The result for a scalar particle
might lead one to expect the path integral propagator $K$ to be identical to
the
field theory one. This turns out not to be the case. Massless chiral
superfield theory
is described by the propagator
\be
\langle{\bar\Phi(x_f,\theta_f,\bar\theta_f)\Phi(x_i,\theta_i,\bar\theta_i)}
\rangle
=i\D^2_f\bar\D^2_f
\left\{{
\int{
{{d^3p}\over{(2\pi)^2}}
{{e^{ip(x_f-x_i)}}\over{p^2}}\delta^4(\theta_f-\theta_i)}}\right\}.
\label{chiral}
\ee
(It is customary to associate the derivatives with the vertices, but this is
only a matter of book keeping.) The derivative structure is the same, the
momentum dependence is not. It is however not a priori clear what
$\langle{\bar\Phi\Phi}\rangle$ means in terms of  propagating from one point in
superspace to another, or, indeed, what that means in more physical terms.
To make
a meaningful comparison between the path integral and superfield theory, we
have
to consider quantities where we have better a priori  reasons for thinking that
the result should be the same. For this reason we turn to the calculation of
the effective action induced by a superparticle in an external gauge field.
As far as the particle is concerned we are then considering a closed loop
and the question of the  physical meaning of moving between points in
superspace disappears.

\bigskip
\begin{flushleft}
{\bf V.  A Supergraph Survey}
\end{flushleft}
\bigskip

As we saw in the previous section, our construction of the path integral for
the superparticle does not result in the propagator that is ordinarily used
in supergraph calculations. We had earlier noted that the superparticle
couples to an external gauge field through the dimensionful gauge
potentials. Superfields, on the other hand, couple through the dimensionless
prepotential. It is conceivable that the differences in coupling and
propagator in the two descriptions cancel and that the particle path integral
and superfield description give the same results for physical quantities. In
this section we review the superfield results we need to make a comparison,
in particular we describe a way of rewriting the supergraph rules such that
gauge potentials  rather than prepotentials do enter the calculation. For
further information about these topics we refer the reader to \cite{BOOK}.

The superfield action for a massless chiral field coupled to an external
electromagnetic gauge field is
\be
S=\int{d^3xd^2\theta d^2\bar\theta \bar\Phi e^{-2V}\Phi},\label{moan}
\ee
where $V$ is a dimensionless prepotential. Gauge transformations act on the
fields as follows:
\be
\Phi &\to &e^{i\Lambda}\Phi, \quad \bar\D\Lambda =0,\cr
\bar\Phi&\to &\bar\Phi e^{-i\bar\Lambda}, \quad \D\bar \Lambda =0,\cr
e^{-2V}&\to & e^{i\bar\Lambda}e^{-2V}e^{-i\Lambda}.
\ee
The supergraph rules are obtained by expanding the action (\ref{moan}) in
powers
of $V$ to get the vertices, and by tying these together by the
$\langle{\bar\Phi\Phi}\rangle$ propagator given in (\ref{chiral}). The
effective
action is the sum of one loop graphs. We will not discuss the details of the
calculation, but remind the reader that at the end of the day the result can be
expressed in terms of gauge potentials and field strengths.
This property can be made more manifest by reorganizing the calculation in a
way
similar to the "doubling" trick in $QED$. The essential idea is that
the result of the one-loop calculation is the superdeterminant of the
equations of motion operator. By considering the square of this operator
instead  one can show that the effective action $\Gamma$ can be written, (the
notation $\D^2\bar\D^2$, et c., was introduced below (\ref{Ddelta})),
\be
e^{i\Gamma}&=& exp\left\{{-i\int{d^3xd^2\bar\theta\half {\delta \over {\delta
j}}\D^2\left({\bar\nabla^2-\bar\D^2}\right)
{\delta \over {\delta
j}}}}\right\}\cr
&&\times exp\left\{{\ihalf \int{d^3xd^2\bar\theta j\partial
^{-2}j}}\right\}\mid _{j=0}, \ee
where
\be
\bar\nabla _\alpha &\equiv & \bar\D_\alpha -2(\bar\D_\alpha V)\cr
{{\delta j(x,\bar\theta )} \over {\delta j(x',\bar\theta ')}}&=&{\cal
D}^2\delta^3(x-x')\delta ^4(\theta-\theta ').
\ee
The result is that the effective action is obtained by evaluating graphs
consisting of closed strings
\be
...\left[{-i\D^2(\bar\nabla^2-\bar{\cal
D}^2)}\right]_{k}\left({{{-i\delta^4(\theta _k-\theta _{k-1})}\over
{p^2}}}\right)\left[{-i\D^2(\bar\nabla^2-\bar{\cal
D}^2)}\right]_{k-1}....
\ee
At this point we depart from the standard treatment and note that once vertex
$k$ has operated on $\delta ^4(\theta _{k}-\theta_{k-1})$, the
expression  is chiral as a function of $\theta _{k}$. We can thus, without
changing anything, insert the chiral projection operator and thus replace the
propagator ${{{-i\delta^4(\theta _k-\theta _{k-1})}\over
{p^2}}}$ by the propagator we derived from the path integral. This can be
done for all propagators in the loop. Thus we have shown that the effective
action corresponding to the coupling of a chiral field to an external $U(1)$
gauge field can be calculated  using our path integral derived propagator and
using $i\D^2(\bar\nabla^2-\bar{\cal
D}^2)$ as vertex. In the next section we will show that something
equivalent to this vertex can be derived from the particle path integral
point of view, and thus show that the particle path integral and the field
theory calculations indeed give the same result.
\bigskip
\begin{flushleft}
{\bf VI.  First Quantized Feynman Rules}
\end{flushleft}
\bigskip

In the previous section we showed that superfield supergraph calculations
can be organized in such a way that the propagator becomes the one we
derived by path integral quantization of a superparticle. This section will
be devoted to the vertex.

The contribution to the effective action of the $k$-th vertex is, according
to the
superfield theory, given by
\be
\int{d^3x_kd^2\theta _kd^2 \bar\theta _k K_0(k+1;k){{(-i)}\over 2} \D^\alpha _k
\D^\beta _k\left({\bar\nabla^k _\beta \bar\nabla^k _\alpha -\bar\D^k _\beta
\bar\D^k _\alpha}\right)K_0(k;k-1)},
\ee
where $K_0$ is the free propagator
\be
K_0(f;i)=-\ihalf \D^\alpha _f\D^\beta _f \bar\D^i _\beta \bar \D^i _\alpha
\left[{
\int{
{{d^3p}\over{(2\pi )^3}}{{\delta^4(\theta _f-\theta
_i)}\over{p^4}}e^{ip\cdot (x_f-x_i)}}
}\right].
\ee
This can equivalently be written as
\be
&&\int{
d^3x_kd^2\theta _kd^2 \bar\theta _k K_0(k+1;k)i\left\{{
2i {\buildrel
\leftharpoonup \over \partial}^k _\mu (\Gamma ^\mu )^\beta _\alpha
\left({
\D^\alpha \bar\D_\beta V
}\right)
}\right.
}\cr
&&\left.{+\left({\D^\alpha \D^\beta \bar\D _\beta \bar\D _\alpha
V}\right) -2 \D^\alpha \D^\beta \left({\bar\D _\beta V \bar\D _\alpha
V}\right)}\right\}K_0(k,k-1),
\ee
where we have integrated by parts and made use of the chirality properties
of $K_0$ and the definition of $\bar \nabla _\alpha$. Note that the vertex
is asymmetric, as was the propagator, and note also that it contains both
terms linear in $V$ and quadratic in $V$ ("seagull" terms). The interaction
lagrangian (\ref{iact}) contains only part of the quadratic term. We will
construct the discretized version of the interaction in such a  way that it
gives the linear term. That second order perturbation theory automatically
reproduces the "seagull" terms is then a nontrivial check on the
construction. Ultimately this goes reflects the gauge invariance, of course.

The problem of discretizing the path integral has two parts: The action and
the measure. We begin with the latter. Recall that for the free particle the
measure is given by
\be
\prod_{k=1}^{N-1} {{d^3x_k d^3p_k d^2\theta _k d^2\bar\theta_k}
\over{(2\pi )^3(p_k^2+m^2)}}
{{d^3p_N}
\over{(2\pi )^3(p_N^2+m^2)}}.
\ee
When interactions are present the mass enters in the action in the
combination $m-\D^\alpha \bar\D_\alpha V$. This suggests that the measure
should be given by
\be
\prod_{k=1}^{N-1} {{d^3x_k d^3p_k d^2\theta _k d^2\bar\theta_k}
\over{(2\pi )^3(p_k^2+(\D^\alpha\bar\D_\alpha V)^2_{k-1})}}
{{d^3p_N}
\over{(2\pi )^3(p_N^2+(\D^\alpha\bar\D_\alpha V)^2_{N-1})}}.
\ee
Here, as well as in what follows, we have restricted ourselves to the $m=0$
case.

Next we turn to the discretization of the interaction part of the action. In
the corresponding treatment of a non-relativistic particle in an external
field this is the place where the midpoint rule (\ref{midp}) plays an
important role. With the midpoint rule a discretized action exists which is
consistent with gauge invariance and has the correct naive continuum limit,
without
it one has to sacifice the continuum limit to preserve gauge invariance. We
already
noted that the vertex we want to recover looks asymmetric and so does the
propagator we constructed. With hindsight we sacrifice the midpoint
construction
and pay the price. We suggest the following discretization:
\be
S&=&\sum\limits_{i} \varepsilon \left\{{ \pi ^\mu _{i+1,i} (\Gamma _\mu)^\beta
_\alpha \left({\D^\alpha \bar\D _\beta
V}\right)_i-(\D^\alpha\bar\D_\alpha V)^2_i}\right.\cr &&\left.{
+2i{{\left({\bar\theta _{i+1}-\bar\theta _i}\right)^\alpha }\over \varepsilon
}(\bar\D_\alpha V)_i +\left({\D^\beta \D^\alpha \bar\D_\alpha
\bar\D_\beta V}\right)_i}\right\}.\label{Idis}
\ee
Before showing that this reproduces the linear term of the vertex, we
discuss the continuum limit and gauge invariance.

The naive continuum limit is
\be
S=\int _{0} ^{T}{
dt\left\{{
\pis ^\beta _\alpha
\D^\alpha \bar\D_\beta V+2i\dot{\bar\theta}^\alpha\bar\D_\alpha V
-(\D^\alpha\bar\D_\alpha V)^2
+\left({
\D^\beta\D^\alpha\bar\D_\alpha\bar\D_\beta V
}\right)
}\right\}}.
\ee
The first three terms are reasonable, they just differ from the interaction
terms in (\ref{sup1}) by a total derivative. The last term is the price we
pay for the asymmetry. If we make a gauge
transformation (\ref{Vgtf}), ${\textstyle{\delta V=i(\bar \Lambda
-\Lambda)}}$, the result should be a function of the endpoints. We obtain
\be
\delta S=\int_{0}^{T}{dt\left\{{-2\pi ^\mu \partial _\mu \bar\Lambda -
2\dot{\bar \theta}^\alpha \bar\D _\alpha \bar\Lambda-2i \partial
^2\bar\Lambda}\right\}},
\ee
reminiscent
of the Ito formula \cite{SCHU}
\be
f(b)-f(a)=\int _{a} ^{b}{dt{{dx}\over{dt}} {{df}\over{dx}}}+\int
 _{a} ^{b}{dt\half\left({{{d^2f}\over{dx^2}}}\right)},
\ee
an equation which is valid in a (Wiener) path integral if the integrals in
the equation are understood as asymmetric Riemann sums:
\be
\int _{a} ^{b}{
dt{{dx}\over{dt}} {{df}\over{dx}}
}
=\mathop{lim}\limits_{N\to \infty}\sum\limits_{k=1}^{N}
{\varepsilon
{{(x_k-x_{k-1})}\over \varepsilon}
{{df}\over{dx}}\left({x_{k-1}}\right)
}.
\ee

After this discussion of the discretized action, we now proceed to show
that it does reproduce the linear terms of the field theory vertex. The path
integral for the effective action is given by
\be
2i\Gamma =\mathop{lim}\limits_{N\to \infty} \int_{0} ^{\infty}{{{dT}\over
T}}\int{\prod_{k=1}^{N} {{d^3x_k d^3p_k d^2\theta _k d^2\bar\theta_k}
\over{(2\pi )^3(p_k^2+(\D^\alpha\bar\D_\alpha V)^2_{k-1})}}}e^{i(S_0+S_I)}
\ee
where $S_0$ is the discretized free action (\ref{Fdis}) and $S_I$ is the
discretized interaction (\ref{Idis}). Note that the measure contains
the same number of $(x,\theta,\bar\theta)$ and $p$ integrations, in contrast
to the situation for the propagator. This is because we are calculating a
loop, so we have to integrate over the initial $=$ final point as well. That
we are calculating a loop is also reflected in the $T^{-1}$-factor in the
$T$-integral \cite{POLY}.

Just as for the non-relativistic particle \cite{FEYN}, the effect of the
interaction is reduced to calculations of certain transition elements and we
wish to relate these to derivatives of the propagator. For the linear order
of the $k$th term in the discretized interaction we need to calculate
expressions of the form

\eject

\be
&&\int{
d^3x_{k+1}d^4\theta_{k+1} d^3x_kd^4\theta _kG(x,\theta ,\bar\theta
;k+1)G(k+1;k;\varepsilon )
}\cr
&&i\varepsilon\left\{{\pi^\mu _{k+1,k}(\Gamma _\mu )^\beta _\alpha
\left({\D^\alpha \bar\D_\beta
V}\right)_k+2i{{(\bar\theta_{k+1}-\bar\theta_k)^\alpha}\over\varepsilon
}\left({\bar\D_\alpha
V}\right)_k}\right.\cr
&+&\left.{\left({
\D^\alpha \D^\beta \bar\D_\beta \bar\D _\alpha
V
}\right)_k
}\right\}G(k;y,\eta,\bar\eta ),\label{mess}
\ee
where
$G$ denotes the free propagator (\ref{disc}). Using the explicit expression
(\ref{disc}) for $G(k+1;k;\varepsilon )$ it is easy to show that
\be
G(k+1;k;\varepsilon )\pi^\mu _{k+1,k}&=&2G(k+1;k;\varepsilon) i{\buildrel
\leftharpoonup \over \partial}^\mu
_k-{{2i}\over\varepsilon}\int{{{d^3p}\over{(2\pi
)^3}}{{p^\mu}\over{p^4}}e^{i\varepsilon\left\{{p\cdot
\pi_{k+1,k}-p^2}\right\}}}\cr
G(k+1;k;\varepsilon)
{{(\bar\theta_{k+1}-\bar\theta_k)^\alpha}\over\varepsilon}&=&{{-1}\over
\varepsilon}\D^\beta _k\int{
{{d^3p}\over{(2\pi)^3}}{{\ps^\alpha _\beta}\over
{p^4}}e^{i\varepsilon\left\{{p\cdot \pi_{k+1,k}-p^2}\right\}
}
}.\label{pirel}
\ee
We insert these expressions in (\ref{mess}) and integrate the $\D^\beta _k$
term by parts. Since $G(k;y,\eta,\bar\eta )$ is antichiral, only the term
where  $\D^\beta$ acts on  $\bar\D_\alpha V$ survives and cancels the second
term in the expression for $G(k+1;k;\varepsilon )\pi ^\mu$. The factor
$\varepsilon$ becomes the measure in the integral over the time at which the
integration acts, and thus the net result is
\be
\int{
dt\int{
d^3x_kd^2\theta_kd^2\bar\theta_k}}&&G(x,\theta,\bar\theta;k)i
\left[{
2i\buildrel\leftharpoonup \over \partial ^\mu
_k\left({
\Gamma_\mu
}\right)^\beta _\alpha
\left({
\D^\alpha \bar\D_\beta
V
}\right)
}\right.
\cr
&&+\left.{\left({
\D^\alpha\D^\beta\bar\D_\beta\bar\D_\alpha
V
}\right)
}\right]G(k;y,\eta,\bar\eta ).
\ee
We have reproduced the linear part of the vertex. In the above expression the
propagators are $G$ rather than $K_0$, but this difference disappears once we
integrate over $T$ and the times at which the interactions occur: The particle
path integral gives us the parametric representation of amplitudes. Before we
proceed with a discussion of the "seagull" terms we emphasize that the
discontinuous
character of the paths in superspace make objects like $\pi ^\mu$ and
$\dot{\bar\theta}$ divergent. It is only the sum of the $\pi^\mu A_\mu$ and
$\dot{\bar\theta}^\alpha \bar A_\alpha$ terms that is finite. Note also that
the
antichiral character of the propagator was important for the cancellation.

The quadratic terms in the vertex arise in second order perturbation theory.
We get contributions both from expanding ${\textstyle{e^{iS_I}}}$ and from
the measure. From the former we get quadratic terms of the form {\it (vertex at
j)$\times$(vertex at k)}, $j\not =k$. These are treated as above. In addition
we get a "contact" term
\be &&{{(i\varepsilon )^2}\over{2!}}\int{
d^3x_{k+1}d^4\theta_{k+1} d^3x_kd^4\theta _kG(x,\theta ,\bar\theta
;k+1)G(k+1;k;\varepsilon )
}\cr
&&\left\{{\pi^\mu _{k+1,k}(\Gamma _\mu )^\beta _\alpha
\left({\D^\alpha \bar\D_\beta
V}\right)_k+2i{{(\bar\theta_{k+1}-\bar\theta_k)^\alpha}\over\varepsilon
}\left({\bar\D_\alpha
V}\right)_k}\right.\cr
&+& \left.{\left({
\D^\alpha \D^\beta \bar\D_\beta \bar\D _\alpha
V
}\right)_k
}\right\}^2G(k;y,\eta,\bar\eta ).\label{mess2}
\ee
For the terms containing $\D^2\bar\D^2V$ we can use the same trick as for
the linear terms in the vertex. $\D^2\bar\D^2V$ is antichiral, so the
integration by parts is not affected and the result for these terms is
\be
{{(i\varepsilon)^2}\over{2!}}{
\int{
d^3x_kd^2\theta_kd^2\bar\theta_k}}&&G(x,\theta,\bar\theta;k)
\left[{
2\cdot 2i\buildrel\leftharpoonup \over \partial ^\mu
_k\left({
\Gamma_\mu
}\right)^\beta _\alpha
\left({
\D^\alpha \bar\D_\beta
V
}\right)_k\left({
\D^\rho\D^\sigma\bar\D_\sigma\bar\D_\rho
V
}\right)_k
}\right.
\cr
&&+\left.{\left({
\D^\alpha\D^\beta\bar\D_\beta\bar\D_\alpha
V
}\right)_k^2
}\right]G(k;y,\eta,\bar\eta )\label{equad}
\ee
The rest of the calculation is no different in principle from the linear
calculation, but messier. One writes down expressions for $\pis _{k+1,k
\alpha} ^\beta \pis _{k+1,k
\rho} ^\delta $ et.c., similar to (\ref{pirel}) and integrates by
parts. Just as for the non-relativistic particle, one gets some extra terms in
addition to the square of the linear contribution. The total result,including
the terms in (\ref{equad}), is
\be
&&{{i^2}\over {2!}}\varepsilon^2
\int{
d^3x_kd^2\theta_kd^2\bar\theta_k}G(x,\theta,\bar\theta;k)
\left\{{
\left[{
\left({
{2i\buildrel\leftharpoonup \over \dels }
}\right)^\beta _\alpha
\left({
{2i\buildrel\leftharpoonup \over \dels }
}\right)^\delta _\gamma
\left({
\D^\alpha \bar\D_\beta
V
}\right)_k
\left({
\D^\gamma \bar\D_\delta
V
}\right)_k
}\right.
}\right.
\cr
&&+\left.{
\left.{
2\left({
{2i\buildrel\leftharpoonup \over \dels }
}\right)^\beta _\alpha
\left({
\D^\alpha \bar\D_\beta
V
}\right)_k
\left({
\D^\gamma\D^\delta\bar\D_\delta\bar\D_\gamma
V
}\right)_k
+\left({
\D^\alpha\D^\beta\bar\D_\beta\bar\D_\alpha
V
}\right)_k ^2
}\right]
}\right.\cr
&&\left.{
+{{2i}\over \varepsilon}\left[{2\D^\alpha\D^\beta(\bar\D_\beta
V\bar\D_\alpha V)_k-(\D^\alpha\bar\D_\alpha V)^2_k}\right]
}\right\}
G(k;y,\eta,\bar\eta )\cr
&&+\int{
d^3x_{k+1}d^4\theta_{k+1}d^3x_kd^4\theta_k{{d^3p}\over{(2\pi
)^3}}G(x,\theta,\bar\theta;k+1)}\cr
&&\times e^{i\varepsilon (p\cdot \pi _{k+1,k}-p^2)}{{\left({
\D^\alpha
\bar\D_\alpha V
}\right)^2}\over{p^4}} G(k;y,\eta , \bar\eta ).
\ee
The ${\cal O}(\varepsilon ^2)$-terms combine with the ({\it  vertex
j})$\times$({\it vertex k}) terms to give the second order contribution of
the linear terms in the field theory vertex. ($\varepsilon ^2$ is the
measure for integration over two independent interaction times.) The $\cal
O(\varepsilon)$ term, combined with the $(\D\bar\D V)^2$ term in the
action, reproduces the field theory "seagull" terms and the ${\cal
O}(\varepsilon ^0)$ term cancels the contribution from the measure to this
order. This completes the demonstration that the second order perturbation
theory based on the discretized interaction (\ref{Idis}) coincides with the
superfield theory. The full equivalence is shown in the appendix .

\bigskip
\begin{flushleft}
{\bf VII.  Conclusions}
\end{flushleft}
\bigskip

As mentioned in the introduction, the massive Siegel invariant superparticle
has
been quantized (in various dimensions) using BRST-methods \cite{GREE}, as well
as
canonical ones \cite{EVAN1}, (without calculating the propagator however). No
difficulties of principle were encountered. For the massless case, the
canonical
procedure is faced with the difficulty that a covariant separation of first and
second class constraints is impossible \cite{BENG}. This has led to
quantization using Batalin-Vilkovisky type Lagrangian BRST methods
\cite{LIND1}.
These constructions involve an infinite tower of ghosts. Our construction would
seem to circumvent both these difficulties. We have ignored the Siegel
invariance, although the (anti)chirality with respect to the endpoints may be
viewed as a remnant of this symmetry. The issue of first and second class
constraints never arises, the construction involves no ghosts, and the limit $m
\to 0$ seems unproblematic. Our treatment of the path integral may seem to be
particular to $D=3, N=2$, but it is clear that this construction of the
propagator works just as well for the $D=4, N=1$ massless superparticle. What
about other cases?

The assumptions in this
article are, (explicitly), the composition property (\ref{cmp}) and,
(implicitly), that the exponent in the infinitesimal propagator should look
"reasonable" as a discretization of the action. In a conventional treatment of
the path integral with gauge fixing and ghosts, a modified version of
(\ref{cmp})
involving also ghost coordinates should be satisfied, since this is essentially
the completeness property of intermediate states. Truncating to $(x^\mu,\theta,
\bar \theta)$ is thus an assumption that the ghost coordinates decouple, which
is reasonable. The second assumption is more questionable. Gauge fixing would
certainly modify the action and with the gauge condition $\dot{\theta}=0$
\cite{SIEG2}, the $\theta$-dependent part of the propagator would be $\delta
^N(\theta _f-\theta _i)$, (which certainly satisfies the composition property).
The role of the ghosts is to remove unphysical states from this, i.e., to
project onto an irreducible representation of supersymmetry. This is precisely
what the projection operator in our construction accomplishes. (C.f. the
situation in string theory with ghosts versus the Brink-Olive projection
operator \cite{GSW}). It thus seems that the existence of a projection operator
that projects onto an irreducible representation of supersymmetry is the basic
requirement in our construction.

We also want to draw attention to an alternative way of viewing our results.
Starting
from the three basic ingredients, the dynamical relation $i\partial _\tau
G=((-i\partial _x)^2 +m^2)G$, the composition property (\ref{cmp}) and
supersymmetry,
we find a limited number of possible definitions of the path integral for a
supersymmetric object satisfying $p^2+m^2=0$. All involve projection operators
acting on $\delta (\theta _f-\theta _i)$. If we confine ourselves to the usual
independent set of projection operators \cite {BOOK}, they give path integrals
which
can be viewed as arising from the usual superparticle action. In this sense we
{\it derive} that action from the path integral!

\bigskip

{\bf Note added:} After the completion of this work we became aware of a
related paper
\cite{FRAD}. Starting from the superfield theory, these authors derive a path
itegral
representation of the Greens functions which involves the superparticle action.
They
give no explicit definition of the path integral, however.

\bigskip

\begin{flushleft}
{\bf Acknowledgements:} We thank H.Hansson, A.Karlhede, M.Ro\v cek and W.Siegel
for numerous discussions on the subject of this article. We also thank
R.Marnelius
for drawing our attention to \cite{FRAD}
\end{flushleft}

\eject

\bigskip
\begin{flushleft}
{\bf Appendix}
\end{flushleft}
\bigskip

In this appendix we prove that the equivalence between superfield theory and
superparticle path integral that we demonstrated to second order in Section
${\bf VI}$
holds to all orders.

Consider the contribution of the $k$-th term in the discretized free action
(\ref{Fdis}) and the interaction (\ref{Idis}). It is given by
$$\int{
{{d^3p}\over {(2\pi )^3}}d^3x_kd^4\theta_k
{{e^{i\varepsilon (p\cdot\pi _{k+1,k}-p^2)}}\over{p^2+(\D^\alpha \bar\D_\alpha
V)^2_k}}}\times$$
$$exp \left\{{
i\varepsilon \left[{ \pi ^\mu _{k+1,k} (\Gamma _\mu)^\beta
_\alpha \left({\D^\alpha \bar\D _\beta
V}\right)_k-(\D^\alpha\bar\D_\alpha V)^2_k}\right.}\right.$$
$$\left.{\left.{
+2i{{\left({\bar\theta _{k+1}-\bar\theta _k}\right)^\alpha }\over \varepsilon
}(\bar\D_\alpha V)_k +\left({\D^\beta \D^\alpha \bar\D_\alpha
\bar\D_\beta V}\right)_k}\right]}\right\}F(x_k,\theta_k,\bar\theta_k),
\eqno{(A1)}$$
where $F(k)$ is the contribution from the terms
$1,...,k-1$. The entire $k$-dependence
of $F$ is carried by $\pi_{k,k-1}$, and thus
$$ \D^\alpha F(k) =0\eqno{(A2)}$$
Expressing $\pi_{k+1,k}$ and $\bar\theta _{k+1}-\bar\theta _k$ in the
interaction as suitable derivatives we can rewrite (A1) as follows:
$$\int{
{{d^3p}\over {(2\pi )^3}}d^3x_kd^4\theta_k F(k)
{
{exp{\left\{{
i\varepsilon
\left[{
-p^2-(\D^\alpha\bar\D_\alpha V)^2_k
+(\D^\beta \D^\alpha \bar\D_\alpha
\bar\D_\beta V)_k
}\right]}\right\}
}
}
\over{p^2+(\D^\alpha \bar\D_\alpha
V)^2_k}
}
}\times$$
$$exp\left\{{
i\varepsilon {1\over{i\varepsilon}}(\D\Gamma ^\mu\bar\D)_k{\partial \over
{\partial p^\mu}}
}\right\} \hat{\cal O}e^{i\varepsilon p\cdot \pi _{k+1,k}}, \eqno{(A3)}$$
where
$$\hat{\cal O}=1-2{{\ps_\alpha^\beta}\over{p^2}}(\bar\D_\beta V)_k\D^\alpha_k
+2{{\ps_\alpha^\beta}\over{p^2}}(\bar\D_\beta V)_k
{{\ps_\gamma^\delta}\over{p^2}}(\bar\D_\delta V)_k\D_k^\gamma\D^\alpha_k.
\eqno{(A4)}$$
Integrating ${\partial \over
{\partial p}}$ in (A3) by parts gives an operator that shifts the variable $p$
by
$(\D\Gamma\bar\D)$, (alternatively, we could have made the shift already in
(A1)).
Inserting also $1=\hat{\cal O}\hat{\cal O}^{-1}$, we get
$$\int{
{{d^3p}\over {(2\pi )^3}}d^3x_kd^4\theta_k F(k)\hat{\cal O}\hat{\cal O}^{-1}
{1\over A}exp{\left\{{-i\varepsilon B}\right\}}\hat{\cal O}
e^{i\varepsilon p\cdot \pi _{k+1,k}}
}, \eqno{(A5)}$$
where
$$A\equiv (p-(\D\Gamma\bar\D V)_k)^2+(\D^\alpha \bar\D_\alpha V)_k^2$$
$$=p^2+2\D^\alpha \D^\beta(\bar\D_\beta V\bar \D_\alpha V)_k +4(\bar\D_\alpha
V)(\D^\alpha \D^\beta \bar\D_\beta V)_k-2(\D\ps\bar\D V)_k$$
$$B\equiv A-(\D^\alpha\D^\beta\bar\D_\beta\bar\D_\alpha V)_k. \eqno{(A6)}$$
We will in fact never need an explicit expression for $\hat{\cal O}^{-1}$,
but it is
clear that it can be constructed in perturbation theory. Next we use the
identity
$$\hat{\cal O}^{-1}f(A,B)\hat{\cal O}=
f(\hat{\cal O}^{-1}A\hat{\cal O},\hat{\cal O}^{-1}B\hat{\cal O}), \eqno{(A7)}$$
and note that antichiral objects commute with $\hat{\cal O}$. Thus
$$\hat{\cal O}^{-1}A\hat{\cal O}=A+\hat{\cal O}^{-1}[A,\hat{\cal O}]$$
$$=A+\hat{\cal O}^{-1}[-2(\D\ps\bar\D V)_k+4(\bar\D_\alpha V
)_k(\D^\alpha\D^\beta\bar\D_\beta V)_k,\hat{\cal O}]$$
$$\hat{\cal O}^{-1}B\hat{\cal O}=\hat{\cal O}^{-1}A\hat{\cal O}
-(\D^\alpha\D^\beta\bar\D_\beta\bar\D_\alpha V)_k.\eqno{(A8)}$$
What saves us from the need of knowing $\hat{\cal O}^{-1}$ is the fact that
$$[-2(\D\ps\bar\D V)_k+4(\bar\D_\alpha V
)_k(\D^\alpha\D^\beta\bar\D_\beta V)_k,\hat{\cal O}]=-\hat{\cal O}4
(\bar\D_\alpha V
)_k(\D^\alpha\D^\beta\bar\D_\beta V)_k.\eqno{(A9)}$$
{\it Proof}: The {\it RHS} can be written as
$$[4(\bar\D_\alpha V
)_k(\D^\alpha\D^\beta\bar\D_\beta V)_k,\hat{\cal O}]-4(\bar\D_\alpha V
)_k(\D^\alpha\D^\beta\bar\D_\beta V)_k\hat{\cal O},\eqno{(A10)}$$
so we only have to show that
$$[2(\D\ps\bar\D V)_k+4(\bar\D_\alpha V
)_k(\D^\alpha\D^\beta\bar\D_\beta V)_k,\hat{\cal O}]=-4(\bar\D_\alpha V
)_k(\D^\alpha\D^\beta\bar\D_\beta V)_k\hat{\cal O}.\eqno{(A11)}$$
which is trivial to verify.

Note that the second derivative term in $\hat {\cal O}$ doesn't contribute to
the {\it
RHS} because it is multiplied by $(\bar \D V)^3=0$.

Returning to (A8) and using (A6), we see that (A5) can be written
$$\int{
{{d^3p}\over {(2\pi )^3}}
d^3x_kd^4\theta_k F(k)\hat{\cal O}}\times$$
$${{exp{\left\{{-i\varepsilon
\left({
-p\cdot\pi_{k+1,k}+p^2-2(\D\ps\bar\D V)_k+2\D^\alpha\D^\beta(\bar\D_\beta
V\bar\D_\alpha
V)_k -(\D^\alpha\D^\beta\bar\D_\beta\bar\D_\alpha V)_k
}\right)
}\right\}
}}
\over
{p^2-2(\D\ps\bar\D V)_k+2\D^\alpha\D^\beta(\bar\D_\beta V\bar\D_\alpha
V)_k}}.\eqno{(A12)}$$
Finally we integrate the $\D$'s in $\hat{\cal O}$ by parts. Using that $F(k)$
is
antichiral, we find that the result replaces the denominator in (A12) by $p^2$.
The
final result is thus
$$\int{
{{d^3p}\over {(2\pi )^3}}
d^3x_kd^4\theta_k}{e^{i\varepsilon (p\cdot\pi_{k+1,k}-p^2)}\over {p^2}}
exp\left\{{
i\varepsilon
\left[{
2(\D\ps\bar\D V)_k-
}\right.
}\right.$$
$$\left.{\left.{
-2\D^\alpha \D^\beta (\bar \D_\beta V\bar\D_\alpha
V)_k+(\D^\alpha\D^\beta\bar\D_\beta\bar\D_\alpha V)_k
}\right]
}\right\}
F(k).\eqno{(A13)}$$
This shows shows agreement with the superfield theory to all orders in
perturbation
theory, (c.f. (64)).

\eject


\begin{thebibliography}{99}
%\bibitem{GREE}
%{D.Amati,} {M. Ciafaloni,} and G. Veneziano.,
%\newblock {\it Phys. Lett.}, {\bf 197B} (1987)81;{\it Int. J. Mod. Phys.},
%{\bf A3} (1988)1615;{\it Phys. Lett.}, {\bf 216B} (1989)41.

\bibitem{EVAN1}
J.M. Evans,
\newblock {\it Nucl. Phys.}, {\bf B331} (1990) 711.

\bibitem{SIEG1}
W. Siegel,
\newblock {\it Phys. Lett.}, {\bf 128B} (1983) 397.

\bibitem{GREE}
{M.B. Green} and C.M. Hull,
\newblock Contribution to {\it "Strings 89"}, World Scientific 1989, 478.

\bibitem{BENG}
{I. Bengtsson,} and M. Cederwall,
\newblock {G\"oteborg Preprint} {\bf 84-21} June 1984.

\bibitem{LIND1}
{U.Lindstr\"om,} {M.Ro\v cek}, {W. Siegel,} {P. van Nieuwenhuizen} and
A.E. van de Ven,
\newblock {\it Phys. Lett.} {\bf 224B} (1989) 285.\\
\newblock{U.Lindstr\"om,} {M.Ro\v cek}, {W. Siegel,} {P. van Nieuwenhuizen} and
A.E. van de Ven,
\newblock {\it Phys. Lett.} {\bf 228B} (1989) 53.\\
\newblock {U.Lindstr\"om,} {M.Ro\v cek}, {W. Siegel,} {P. van Nieuwenhuizen}
and
A.E. van de Ven,
\newblock {\it Journ. Math. Phys.} {\bf 31} (1990) 1761.

\bibitem{ROCE}
{M.Ro\v cek}, {W. Siegel,} {P. van Nieuwenhuizen} and A.E. van
de Ven,
\newblock {\it Phys. Lett.} {\bf 227B} (1989) 87.\\

\bibitem{NISSI}
E. Sokatchev,
\newblock {\it Phys. Lett.} {\bf 169B} (1986) 209; {\it Class. Quantum Grav.}
{\bf 4} (1987)237.\\
\newblock {L. Brink,} {M. Hennaux} and C. Teitelboim
\newblock {\it Nucl. Phys.} {\bf B293} (1987) 505.\\
\newblock {E. Nissimov,} {A. Pacheva} and S. Solomon,
\newblock {\it Nucl. Phys.} {\bf B296} (1988) 462; {\bf B299} (1988) 183.\\

\bibitem{EVAN2}
J.M. Evans,
\newblock {\it Class. Quantum Grav.}, {\bf 7} (1990) 699.

\bibitem{MIKO}
{A. Mikovi\'c} and W. Siegel,
\newblock {\it Phys. Lett.} {\bf 240B} (1990) 363.

\bibitem{NORD}
H.Nordstr\"om, "A light cone superparticle-superfield theory relation",(in
preparation).

\bibitem{GRUN}
{J. Grundberg,} {U. Lindstr\"om} and H. Nordstr\"om,
\newblock "Discretization of the Superparticle Path Integral", University of
Stockholm preprint, USITP-92-11, October 1992, hep-th/9211024

\bibitem{SCHU}
L.S. Schulman, {\it"Techniques and Applications of Path Integration"}, J. Wiley
and sons, New York 1981.

\bibitem{FEYN}
{R.P. Feynman} and  A.R. Hibbs,
\newblock {\it "Quantum Mechanics and Path Integrals"},  McGraw-Hill, New
York 1965.

\bibitem{OHNU}
{Y. Ohnuki} and K. Kashiwa,
\newblock {\it Prog. Theor. Phys.} {\bf 60} (1978) 548.

\bibitem{BOOK}
{S.J. Gates,} {M.T. Grisaru} {M. Ro\v cek} and W. Siegel,
\newblock {\it "SUPERSPACE"}, Benjamin/Cummings, Reading 1983.

\bibitem{GAUN}
J.P. Gauntlett,
\newblock {\it Phys. Lett.} {\bf 272B} (1991) 25.

\bibitem{GAUN2}
{J.P. Gauntlett} and C.F. Yastremiz,
\newblock {\it Class. Quantum Grav.} {\bf 7} (1990) 2089.

\bibitem{SIEG2}
W. Siegel,
\newblock Contribution to {\it "Strings 89"}, World Scientific 1989, 211.

\bibitem{GSW}
{M.B. Green}, {J.H. Schwarz} and E. Witten,
\newblock {\it "Superstring Theory"}, Cambridge University Press, Cambridge
1987.

\bibitem{POLY}
A.M. Polyakov,
\newblock {\it "Gauge Fields and Strings"}, Harwood, 1987.

\bibitem{FRAD}
E.S. Fradkin and SH.M. Svartsman,
\newblock {\it Mod. Phys. Lett.} {\bf A6}, (1991) 1977.


 \end{thebibliography}
 \end{document}